# Terahertz Saturable Absorption in Superconducting Metamaterials

George R. Keiser[1,4*], Jingdi Zhang[3], Xiaoguang Zhao[2], Xin Zhang[2*], and Richard D. Averitt[3*]

[1]*Department of Physics, Boston University, Boston, MA, USA*

[2]*Department of Mechanical Engineering, Boston University, Boston, MA, USA*

[3]*Department of Physics, UC-San Diego, San Diego, CA, USA*

[4]*School of Engineering, Brown University, Providence, RI, USA*

*Correspondence may be addressed to: raveritt@ucsd.edu, xinz@bu.edu, or george_keiser@brown.edu.*

**Abstract:** We present a superconducting metamaterial saturable absorber at terahertz frequencies. The absorber consists of an array of split ring resonators (SRRs) etched from a 100nm YBaCu$_3$O$_7$ (YBCO) film. A polyimide spacer layer and gold ground plane are deposited above the SRRs, creating a reflecting perfect absorber. Increasing either the temperature or incident electric field (E) decreases the superconducting condensate density and corresponding kinetic inductance of the SRRs. This alters the impedance matching in the metamaterial, reducing the peak absorption. At low electric fields, the absorption was optimized near 80% at T=10K and decreased to 20% at T=70K. For E=40kV/cm and T=10K, the peak absorption was 70% decreasing to 40% at 200kV/cm, corresponding to a modulation of 43%.

Metamaterials (MMs) are by now a well-known scheme for engineering the optical response of materials [1,2]. Initial research focused on engineering materials with novel refractive index, n($\omega$) and impedance, Z($\omega$) with negative index of refraction[2] and electromagnetic cloaking[3] being the most prominent results. These applications only scratch the surface of MM applications in materials engineering. MMs have potential use in memory materials[4], thermal detectors[5], waveplates[6], and chemical sensing[7], to name a few examples. MM photoexcitation[8], temperature control[9], electrical modulation[10,11], and structural control of intra-unit cell coupling[12-14] combined with MEMS actuation[15-17] have opened the door for tunable and broad bandwidth MM modulators and switches[18-20]. The ability to control intra unit-cell coupling has produced a particularly promising MM device known as a "perfect absorber" (PA) [21-23]. At THz frequencies, PAs comprise some of the most promising applications, providing new methods for THz sensing and detection, helping to fill the "THz gap" [24]. The use of complex material substrates and optical pumping has made dynamically tunable PAs a reality[11,25].

More recently, the development of high intensity THz sources[26] has jump-started research into "nonlinear" THz MMs, i.e. MMs with optical properties dependent on the intensity of incident radiation. At microwave frequencies, nonlinear MMs can be created by introducing (for example) varactors[27] or compressible inclusions[28] into the design. At THz frequencies, such elements are too large to be of use as unit cell inclusions. However, semiconductor inclusions[29] and correlated electron material substrates[30] can be used to produce highly nonlinear MMs. Recent interest has focused using superconducting (SC) thin films as the basis of new nonlinear MM designs[31]. The complex conductivity, $\sigma(\omega)$, of a SC is sensitive to both temperature and electromagnetic fields, allowing for nonlinear MMs that can be dynamically tuned via temperature, magnetic field, or electric current[32-36]. Additionally, the low intrinsic electrical resistance of SCs allows for low loss, compact MM unit cell designs[37].

In this paper, we combine the idea of SC MMs and PAs to create a nonlinear THz device, a resonant saturable absorber. We introduce a superconducting response into the absorber design by etching the SRR array out of a commercially obtained YBCO film of thickness $d_s$=100nm ($T_c \approx$ 75K, substrate $LaAlO_3$(LAO)) using a lithographic wet etching technique.[38] A polyimide spacer layer and gold ground plane are then placed above the SRRs to form a "reflecting" PA.[21] Figure 1a shows a schematic of the device with dimensions while figures 1b and 1c show photos of the mask used for etching and the resulting YBCO SRRs, respectively. Using experimental conductivity measurements of the YBCO film (Fig. 2, discussed more below), we performed CST Microwave Studio simulations and optimized the dimensions of the SRRs such that the resonant absorption is maximized at 0.5THz for T=10K. The SRR side length is L=40μm, unit cell periodicity, P=56μm, linewidth, w=5μm, and the capacitive gap width, g=3μm. To create the ground plane and spacer layer, a 3μm thick polyimide film was spin coated onto a $Si$ substrate. A 10nm Ti adhesive layer and a 200nm layer of gold were then deposited on the polyimide subsequently. Lastly, the gold-coated polyimide film was pulled from the Si substrate and fixed to the SRR array using the MM "tape" concept[39]. A 200nm layer of photoresist (PMMA) was used as an adhesive.

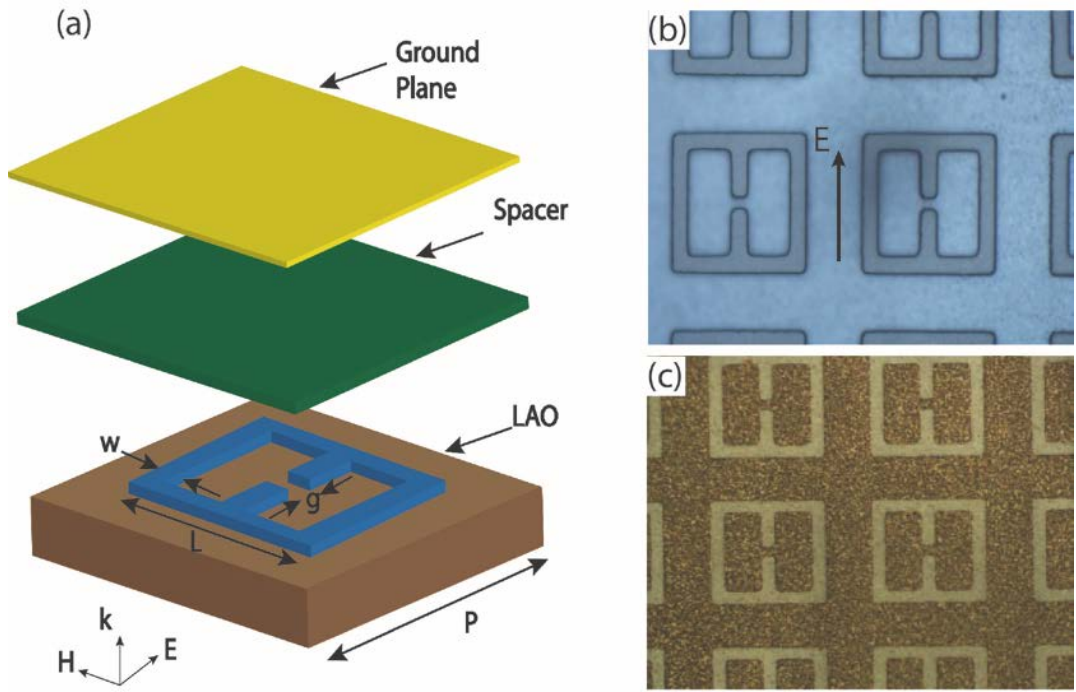

**Figure 1:** (a) Expanded unit cell schematic of a reflecting perfect absorber with relevant dimensions. (b) Mask used to etch YBCO SRRs. (c) Etched YBCO SRRs.

Unity absorption (A) results when a MM's transmittance ($t$) and reflectance (r) are engineered to be simultaneously zero.

$$A = 1 - |t|^2 - |r|^2 = 1 \tag{1}$$

In a reflecting perfect absorber, the gold ground plane reduces t to negligible values. The normal incidence reflectivity, r, from the MM interface is given by[23, 40]:

$$r = \frac{Z-Z_o}{Z+Z_o} \tag{2}$$

where Z is the MM homogenous impedance, and $Z_o$ is the impedance of free space. In the case where Z is matched with $Z_o$, the reflectivity from the interface drops to zero.

Impedance matching is achieved through simultaneous engineering of the MM permittivity, $\epsilon(\omega)$, and permeability, $\mu(\omega)$. THz radiation at normal incidence is polarized so the E field couples to the SRR's fundamental mode, $\omega_o$, via the capacitive gap (Fig. 1a). A resonant $\epsilon(\omega)$ can be tailored by proper choice of the dimensions, L, g, and P[41]. The THz H field simultaneously couples to the MM mode via the gap between the SRR and induced image currents in the ground plane, yielding an effective permeability, $\mu(\omega)$, which can be tuned independently of $\epsilon(\omega)$ by changing the spacer layer thickness, d, until $\mu(\omega_o)=\epsilon(\omega_o)$ and:

$$Z = \sqrt{\frac{\mu(\omega_o)}{\epsilon(\omega_o)}} = 1 \tag{3}$$

The interface reflection thus drops to zero near $\omega_o$ while the high resonant loss in the MM (i.e. $\epsilon_2$) eliminates etalon reflections from the bulk of the MM and thus a resonant unity absorption condition is achieved.

The temperature dependence of the frequency dependent absorption $A(\omega)$ arises from the complex conductivity, $\sigma(\omega)$, of the YBCO film, which is described by a two fluid model.[42] In this model, two populations of charge carriers give rise to the complex conductivity. At THz frequencies, the superfluid condensate contributes an $i/\omega$ term to the conductivity dispersion, while normal state carriers contribute a Drude dispersion. The resulting conductivity has the form:

$$\sigma(\omega, T) = \sigma_o \left[ \frac{f_n(T)}{\tau(\omega,T)^{-1} - i\omega} + f_s(T)\left(\frac{i}{\omega} + \pi\delta(\omega)\right) \right] \tag{4}$$

where $\tau^{-1}$ is the normal carrier scattering rate, and $f_n$ and $f_s$ the relative filling fraction of normal carriers and superconducting condensate, respectively. Below the superconducting transition, changing the ambient temperature modifies the relative filling fractions, giving rise to the temperature dependence.

Figure 2 plots $\sigma(\omega, T)$, for an unetched YBCO film at various temperatures, measured using THz-TDS in transmission. For high temperature, the film conductivity is dominated by a Drude response. As the temperature begins to decrease, Fig. 2c shows that the real part of the conductivity, $\sigma_1$, increases. This is expected from the Drude model since $\tau^{-1}$ decreases for lower temperature. For T=70K and below, the trend shifts and $\sigma_1$ decreases with decreasing T and a $1/\omega$ trend dominates the imaginary component, $\sigma_2$, characteristic of a superconducting state. The onset of this $1/\omega$ response allows for the transition temperature to be estimated at $T_c \sim 75K$. This behavior corresponds well with both the two-fluid model [42], and previously reported data[43].

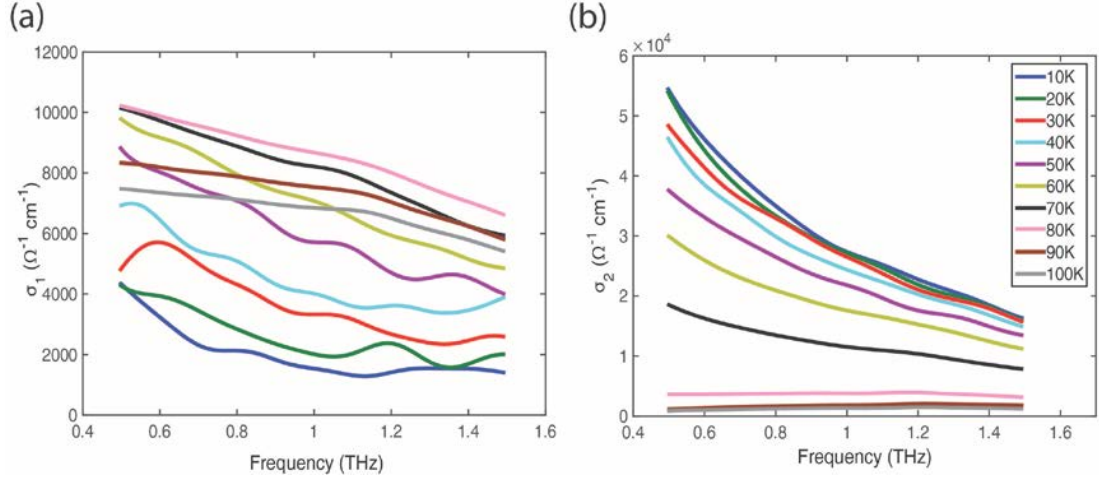

**Figure 2:** (a) Real and (b) imaginary experimental conductivity of the pre-fabrication YBCO for varying temperature.

After fabrication, the YBCO SRRs retain $\sigma(\omega)$. The superconducting fluid affects the SRR resonance through the effective resistance, $R_{eff}$, and the carrier kinetic inductance, $L_k$, which are functions of $\sigma(\omega)$[44, 45]:

$$L_k = \sqrt{\frac{i\mu_o}{\omega\sigma_2(\omega)}} \coth\left(d\sqrt{i\mu_o\omega\sigma_2(\omega)}\right)$$

(5)

$$R_{eff} = \sqrt{\frac{\omega\mu_o}{2\sigma_1(\omega)}}$$

The YBCO SRRs can be treated as lossy LC circuits with resonance frequency, $\omega_o$, given by the eigenmode equation for an RLC resonator:

$$\omega_o = \sqrt{\frac{1}{(L+L_k)C} - \frac{R^2}{4(L+L_k)^2}}$$

(6)

where L is the SRR geometrical inductance and C is the SRR gap capacitance. Increasing temperature will lower $L_k$ and increase $R_{eff}$. The second term in Eq. 6 results from the damping in the system. An increase in the resonator loss broadens the resonance, disrupts impedance matching and decreases the peak absorption.

Saturable absorption with increasing electric field arises from the suppression of the superconducting state through current-induced pair breaking[46, 47]. The intense THz electric fields drive a supercurrent in the SRRs up to and beyond the critical current density, $j_c$. Above, $j_c$, there is a reduction of the superfluid density, raising $R_{eff}$[48], resulting in damping of the resonance with a corresponding decrease of the absorption as described above. In short, the MM absorption depends on the electric field strength of the incident THz radiation as well as the ambient temperature.

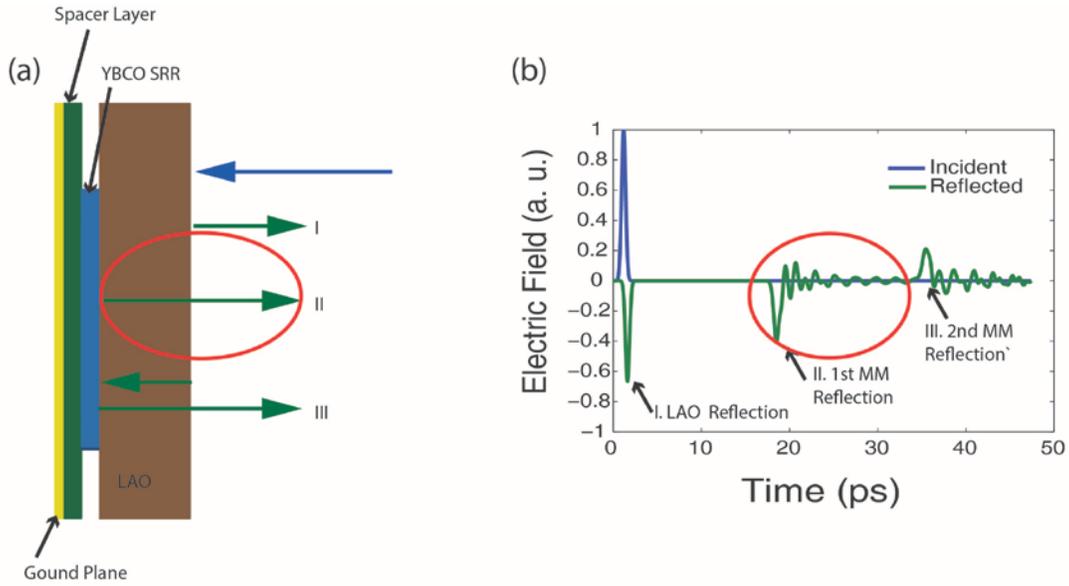

**Figure 3:** Experimental geometry for THz-TDS of the YBCO absorber. (a) Schematic of metamaterial unit cell with etalon reflections. (b) Simulated time domain signals from the absorber showing etalon reflections. The red circles mark the reflection used for this experiment.

High-field THz-TDS in reflection was used to characterize the saturable absorber. THz field strengths up to $E_o$=200kV/cm were generated using the tilted pulse front technique in LiNbO$_3$ [26]. Using two wire grid polarizers to control electric field strength and polarization, pulses between 20 and 200kV/cm were directed onto the absorber and polarized such that the E field coupled to the SRR capacitive gap. The absorber was cooled with liquid He and reflection spectra were measured for different temperatures and electric field strengths. THz transmission through the ground plane is negligible and the absorption spectra are directly obtained from the measured reflectance spectra:

$$A(\omega, T, E) = 1 - r^2(\omega, T, E) \qquad (7)$$

Due to the ground plane layer, direct illumination of the active layer is not possible. Instead, the incident THz beam must first pass through the LAO substrate, as shown schematically in Fig. 3a. The LAO substrate acts as a Fabry-Perot etalon, producing multiple pulses in the time domain reflection signal, shown in Fig. 3a and 3b. The first reflected pulse originates from the LAO/air interface and is of no interest here. The second reflected pulse, circled in Fig. 3b, originates from the MM interface and carries the information about the MM absorption. This second pulse is used to obtain the reflectance spectra. A 500μm LAO substrate with a 3μm polyimide spacer layer and gold ground plane is used as a reference. The reference sample mimics the optical dispersion and path length of the MM, but does not contain the YBCO layer. Thus, the absorption values reported below are normalized to the strength of the signal that is transmitted through the substrate, measuring the *internal absorbance* of the device.

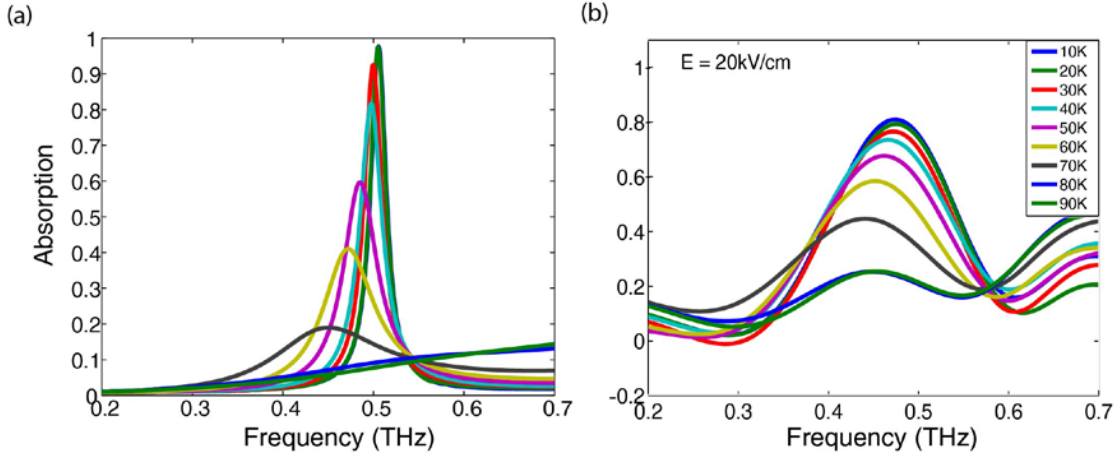

**Figure 4:** (a) Simulated MM absorption spectrum for varying temperature (b) Experimental absorption spectra at E=20kV/cm and varying temperature. The baseline oscillation is an artifact of an etalon reflection in the measurement.

Figure 4b shows the experimental absorption curves for a THz field strength of 20kV/cm at different temperatures. For comparison, Fig. 4a shows the simulated absorption spectra (using the experimentally measured conductivity data from Figure 2). In both simulation and experiment, the absorption peak is maximized for T=10K and decreases with increasing temperature until the peak disappears entirely for T>Tc≈75K.

Multiple factors can increase the resonance width in experiment in comparison to simulation. Fabrication error in the spacer layer thickness and changes in the material properties of the YBCO during fabrication can broaden the absorption. However, for the present experiments, limited spectral resolution as a result of the etalon reflections in the time domain data is the largest contributor. The same etalon produces the baseline modulation in the experimental data. This modulation and the associated points of "negative absorption" are, of course, non-physical artifacts of the experiment.

Both simulation and experiment show a redshift in the absorption peak with increasing temperature, caused by the increase in $R_{eff}$ and corresponding decrease in $L_k$. Here we see the importance of considering the loss in the SRRs. The observed redshift depends on changes to both $L_k$ and $R_{eff}$. Assuming lossless resonators with $R_{eff} \approx 0$ for all T and a fundamental mode given by $1/\sqrt{(L + L_k)C}$, one would predict a blueshift with increasing T (lower $L_k$). Instead, in the lossy SRRs raising temperature causes the second term in Eq. 6, $\frac{R_{eff}^2}{(L+L_k)^2}$, to grow more quickly than the first term, shifting the mode to lower frequencies.

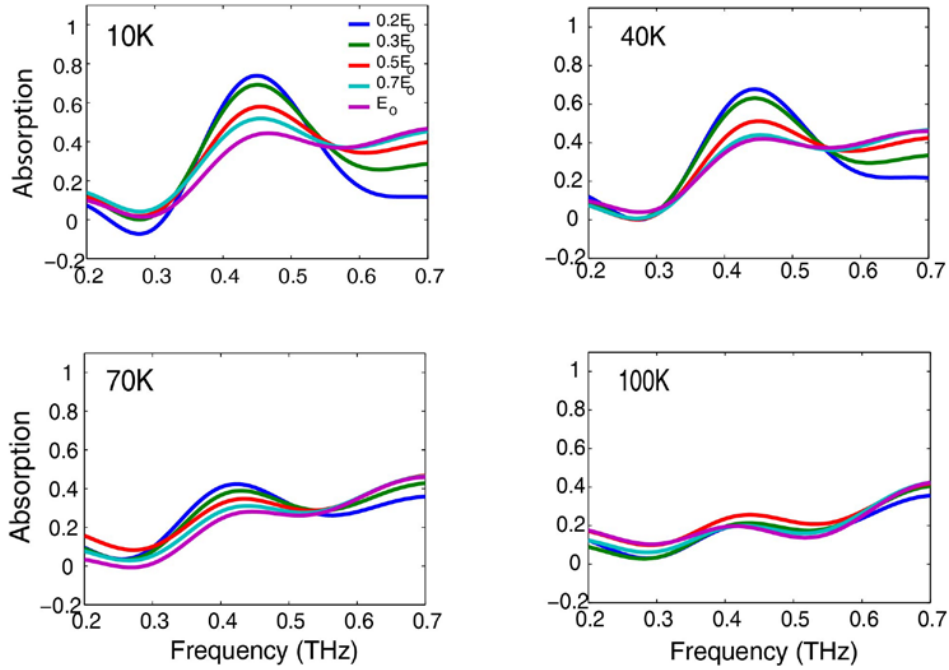

**Figure 5**: Field dependence of the YBCO MM absorption spectrum at 10K, 40K, 70K, and 100K. $E_o$=200kV/cm.

Figure 5 shows the absorption spectra for varying incident field strength at four temperatures. The absorber is strongly sensitive to incident field strength and functions as a saturable absorber for 10K≤T≤70K. The saturation with increasing electric field is most pronounced at low temperatures. At T=10K (Fig. 5a), increasing E from $0.2E_o$ =40kV/cm to $E_o$=200kV/cm causes the MM absorption peak to decrease from 0.70 to 0.40, corresponding to a modulation of 43% of the peak absorption. Though reduced, saturable absorption is present for higher T, up to T=$T_c$. For T=40K (Fig. 5b) and T=$T_c$=70K (Fig. 5c), the saturable absorption effect follows a similar trend, but with less modulation. For T>Tc=100K (Fig. 5d) and higher, the resonance is saturated completely for all field strengths.

A notable feature in Figure 4 is the absence of a redshift in the absorption peak with increasing E field (in contrast to the temperature dependent results in Fig. 4). This has also been seen in a previous work[44] of superconducting SRRs in a non-perfect absorber geometry. There it was argued that the electric field induced normal carriers screen the superconducting film (as a whole) to incident fields. As a result, $L_k$ is effectively unchanged, eliminating the redshift. This explanation appear to be unlikely since $L_k$ originates from the $i/\omega$ term in Eq. 4 and is characteristic of the superconducting state. A decrease in condensate density, induced by either temperature or electric field, must correspond to a decrease in $L_k$ for any superconducting material.

It is also possible that $R_{eff}$ experiences a slower growth and lower peak value as E is increased instead of T. This difference could arise from differences in τ (see eq. 4), for carriers induced by supercurrent pair breaking, vs. those induced by increasing T. Normal state carriers induced via supercurrent sit in a lower temperature environment and thus have a lower τ. $R_{eff}$ would thus be lower for the case of electric-field induced normal carriers, resulting in less of a redshift. However, more work is needed before conclusive remarks can be made on this subject.

To conclude, we combined the ideas of metamaterial perfect absorbers and superconducting metamaterials to create a saturable absorber at terahertz frequencies. The metamaterial consists of an array of split ring resonators (SRRs) etched from a 100nm $YBaCu_3O_7$ (YBCO) film. A polyimide spacer layer and gold ground plane are placed above the SRRs to create a reflecting perfect absorber. Changing ambient temperature or terahertz field strength altered the complex conductivity of the YBCO SRR array and decreased impedance matching in the absorber, reducing the peak absorption. We used numerical simulations and high-field terahertz time domain spectroscopy to characterize the absorber. For an incident terahertz field strength of E=20kV/cm, the absorption was optimized near 80% at T=10K and decreased to 20% at T=70K, the approximate $T_c$ for the YBCO SRRs. For E=40kV/cm and T=10K, the peak absorption was 70%. At E=200kV/cm and T=10K, the absorption saturates to 40%, a total modulation of 43%. The saturable absorption effect is present over a broad temperature range and is tunable via temperature.

We acknowledge support from DOE-Basic Energy Sciences under Grant No. DE-SC0012592 under which the THz measurements and electron acceleration analysis were performed. In addition, we acknowledge the National Science Foundation under Grant No. ECCS-1309835. The authors would like to thank the Boston University Photonics Center for technical support.